\documentclass[twocolumn,aps,prd,eqsecnum,showpacs,preprintnumbers]{revtex4}

\usepackage{graphicx}
\def\apjs{Astrophys.\ J.\ Suppl.\ }
\def\mnras{Mon.\ Not.\ R.\ Astron.\ Soc.\ }
\begin{document}
\title
{Faraday resonance in dynamical bar instability of differentially
  rotating stars}
%
\author{Motoyuki Saijo}
\email{ms1@maths.soton.ac.uk}
%
\affiliation
{School of Mathematics, University of Southampton, 
Southampton SO17 1BJ, United Kingdom}
%
\author{Yasufumi Kojima}
\email{kojima@theo.phys.sci.hiroshima-u.ac.jp}
%
\affiliation
{Department of Physics, Hiroshima University, Higashi-Hiroshima
  739-8526, Japan}
%
\received{7 August 2007}
\revised{11 January 2008}
\accepted{11 February 2008}
%
\begin{abstract}
We investigate the nonlinear behaviour of the dynamically unstable
rotating star for the bar mode by three-dimensional hydrodynamics in
Newtonian gravity.  We find that an oscillation along the rotation
axis is induced throughout the growth of the unstable bar mode, and
that its characteristic frequency is twice as that of the bar mode,
which oscillates mainly along the equatorial plane.  A possibility to
observe Faraday resonance in gravitational waves is demonstrated and
discussed.
\end{abstract}
%
\maketitle
%
\section{Introduction}
\label{sec:intro}
Parametric resonance is widely observed in hydrodynamics, nonlinear
optics, chemical reaction and classical oscillatory systems.  It is
also interesting from the bifurcation theory and the pattern
formation.  The scientific study in the fluid mechanics dates from the
experiments by Faraday in 1831, and is therefore named Faraday
resonance (e.g., \citep{MH90}).  Nonlinear dynamics exhibits mode
interaction of oscillation in different direction, and possibly causes
the resonant growth of a particular mode.  Recently, experimental
studies of Faraday resonance demonstrate that the system of fluid
mechanics \citep{BAFM06} and that of Bose-Einstein condensate
\citep{EAH07} work perfectly.  These agreements in different fields
also suggest that Faraday resonance may also occur in an astrophysical
context.  Quasi-periodic oscillation in gravitational waves from
dynamical/secular instabilities is expected to be excited throughout
rotating core collapse, and may drive resonant growth.

Dynamical bar instability in a rotating equilibrium star takes place
when the ratio $\beta$ ($\equiv T/W$) between rotational kinetic
energy $T$ and the gravitational binding energy $W$ exceeds the
critical value $\beta_{\rm dyn}$ ($\approx 0.27$ for an uniformly
rotating incompressible body in Newtonian gravity
\citep{Chandra69}).  Determining the onset of the dynamical bar-mode
instability, as well as the subsequent evolution of an unstable star,
requires a fully nonlinear hydrodynamic simulation.  Simulations
performed in Newtonian gravity (e.g.,
\citep{TDM95,DGTB,WT,HCS,SHC,HC,TIPD,NCT,LL,Liu02}) have shown that
$\beta_{\rm dyn}$ depends very weakly on the stiffness of the equation
of state.  $\beta_{\rm dyn}$ becomes small for stars with high degree
of differential rotation \citep{TH90,PDD,SKE}.  Simulations in
relativistic gravitation \citep{SBS00,SSBS01,BPMR07} have shown that
$\beta_{\rm dyn}$ decreases when increases the compactness of the
star, indicating that relativistic gravitation enhances the bar-mode
instability.  Recent numerical simulations show that dynamical bar
instability can occur at significantly lower $\beta$ than the
threshold $\beta_{\rm crt} \approx 0.27$
\citep{CNLB01,SKE,SBS03,OT06,CQF07}.  These recent findings can be
classified into the category of a low $T/W$ dynamical
instability.  This instability may be triggered by the corotation
resonance \citep{WAJ05,SY06}, which is completely different from the
standard dynamical bar-mode instability triggered by a certain
magnitude of rotation \citep{Chandra69,Tassoul78,ST83}.

Our main concern in this paper is not to determine the onset of the
instability, but to study the dynamical features of the bar.  For this
purpose, we numerically study the growing behaviour of the azimuthal
modes in the nonlinear regime for a longer timescale.  One interesting
issue of nonlinear evolution is the possibility of resonant growth of
other azimuthal modes triggered by the dynamical bar-mode
instability.  One candidate for such resonance is Faraday resonance,
which is excited by the external periodic force.  According to the
linear approximation of the velocity potential by using an
incompressible inviscid liquid in a rectangular tank, the
time-dependent behaviour of the liquid surface is expressed by the
Mathieu's equation (e.g., \citep{DR81}).  The dynamically unstable bar
mode may work for other azimuthal oscillation modes as an external
periodic force.  The oscillation is not exactly periodic, but rather
quasi-periodic, and may trigger a parametric resonance.

The other interesting issue of nonlinear evolution is the duration of
the bar shape, when it forms.  This is quite important for
gravitational wave detection.  We basically believe that once the
dynamical bar instability takes place, the system generates
quasi-periodic gravitational waves for a period sufficient enough to
be detected in the ground-based gravitational wave detectors.  The
only causes to destruct a bar are dissipative effects such as
viscosity and gravitational radiation.  The typical timescale of such
effects takes place in the secular timescale, which is much longer
than the dynamical one of the system.  Therefore the standard picture
is that the bar can persist in its shape until the secular timescale.
However, recent numerical simulation shows that a bar destructs its
shape in the dynamical timescale \citep{BPMR07}.  The authors argue a
possible cause of the destruction of bar as azimuthal mode coupling.
Although there was in the past a debate between the two numerical
simulations about the persistence of a bar \citep{NCT,Brown}, the
different outcomes were considered as the different accuracy level of
the center of mass condition at that time.  Since there is the only
one group that claims the destruction of bar structure in the
dynamical timescale with a satisfaction of the center of mass
condition, it is worth investigating the destruction of a bar
employing a different computational code.  In order to focus on this
topic, it is sufficient to investigate this topic in three-dimensional
hydrodynamics in Newtonian gravity.

This paper is organized as follows.  In Sec.~\ref{sec:bequation} we
present the basic equations of our hydrodynamic simulation in
Newtonian gravity.  In Sec.~\ref{sec:nr} are discussed the numerical
results of our findings of Faraday resonance.  In
Sec.~\ref{sec:Conclusion} we briefly summarize our
findings. Throughout this paper, we use the geometrized units with
$G=c=1$ \footnote{The speed of light only enters through the
  quadrupole formula of gravitational waves.} and adopt Cartesian
coordinates $(x,y,z)$ with the coordinate time $t$.  Note that Latin
index takes $(x,y,z)$.

\section{Basic equations}
\label{sec:bequation}
\subsection{Newtonian Hydrodynamics}
We construct a three dimensional Newtonian hydrodynamics code assuming
an adiabatic $\Gamma$-law equation of state
\begin{equation}
P = ( \Gamma - 1 ) \rho \varepsilon,
\label{eqn:GammaLaw}
\end{equation}
where $P$ is the pressure, $\Gamma$ the adiabatic index, $\rho$ the
mass density and $\varepsilon$ the specific internal energy density.
For perfect fluids the Newtonian equations of hydrodynamics consist of
the continuity equation
\begin{equation}
\frac{\partial \rho}{\partial t}
+\frac{\partial (\rho v^{i})}{\partial x^{i}} = 0,
\label{eqn:continuity}
\end{equation}
the energy equation
\begin{equation}
\frac{\partial e}{\partial t}+
\frac{\partial (e v^{j})}{\partial x^{j}} = 
- \frac{1}{\Gamma} e^{-(\Gamma-1)} P_{\rm vis} 
\frac{\partial v^{i}}{\partial x^{i}}
,
\end{equation}
and the Euler equation
\begin{equation}
\frac{\partial(\rho v_{i})}{\partial t}
+ \frac{\partial (\rho v_{i} v^{j})}{\partial x^{j}} 
=
- \frac{\partial (P + P_{\rm vis})}{\partial x^{i}}
- \rho \frac{\partial \Phi}{\partial x^{i}}.  
\end{equation}
Here $v^i$ is the fluid velocity and $\Phi$ the gravitational
potential; and $e$ is defined according to
\begin{equation}
e = (\rho \varepsilon)^{1/\Gamma}.
\end{equation}
We compute the artificial viscosity pressure $P_{\rm vis}$ from
\citep{RM}
\begin{equation}
P_{\rm vis} =
\cases{ 
C_{\rm vis} 
\rho (\delta v)^{2},
& for $\delta v \leq 0$;
\cr
0, & for $\delta v \geq 0$,\cr
}
\end{equation}
where $\delta v \equiv 2 \delta x \partial_{i} v^{i}$, $\delta x (=
\Delta x = \Delta y = \Delta z)$ is the local grid spacing and where
we choose the dimensionless parameter $C_{\rm vis} = 2$.  When
evolving the above equations we limit the stepsize $\Delta t$ by an
appropriately chosen Courant condition. 

The gravitational potential is determined by the Poisson equation
\begin{equation}
\triangle \Phi = 4 \pi \rho,
\end{equation}
with the outer boundary condition
\begin{equation}
\Phi = - \frac{M}{r} - \frac{d_{i} x^{i}}{r^{2}} + O(r^{-3}).
\end{equation}
Here $M$ is the total mass
\begin{equation}
M = \int_{V} \rho dx^{3}
\label{eqn:M}
\end{equation}
and $d_i$ is the dipole moment
\begin{equation}
d_{i} = \int_{V} \rho x_{i} dx^{3}.
\end{equation}

\subsection{Initial Data}
As initial data, we construct differentially rotating equilibrium
models with an algorithm based on \citet{Hachisu} and adopt
cylindrical two dimensional coordinate to compute the axisymmetric
equilibrium star.  Individual models are parameterized by the ratio of
the polar to equatorial radius $R_{\rm p}/R_{\rm eq}$, and a parameter
of dimension length $d$ that determines the degree of differential
rotation through
\begin{equation}
\Omega = \frac{j_{0}}{d^{2} + \varpi^{2}}.
\label{eqn:omega}
\end{equation}
Here $\Omega$ is the angular velocity, $j_{0}$ a constant parameter
with units of specific angular momentum, and $\varpi$ the cylindrical
radius.  The parameter $d$ determines the length scale over which
$\Omega$ changes; uniform rotation is achieved in the limit $d
\rightarrow \infty$.  For the construction of initial data we also
assume a polytropic equation of state
\begin{equation}
P = \kappa \rho^{1+1/n},
\end{equation}
where $n=1/(\Gamma-1)$ is the polytropic index and $\kappa$ a
constant.  In the absence of shocks, the polytropic form of the
equation of state is conserved by the $\Gamma$-law equation of state
(Eq.~[\ref{eqn:GammaLaw}]).

We also compute the virial identity, which is identically zero in the
equilibrium star, to show the accuracy level as
\begin{equation}
V_{\rm Nwt} = \frac{|2T_{\rm tot} - W + 3 \Pi|}{W},
\end{equation}
where 
\begin{eqnarray}
T_{\rm tot} &=& \frac{1}{2} \int \rho v^{i} v_{i} d^3 x
,\\
W &=& - \frac{1}{2} \int \rho \Phi d^3 x
,\\
\Pi &=& \int P d^3 x
.
\end{eqnarray}
Note that we have divided the value by a gravitational binding energy
$W$ so that the value $V_{\rm Nwt}$ is regarded as a relative error of
the system.  We summarize our four different rotating equilibrium
stars in Table \ref{tab:equilibrium}.

\subsection{Gravitational Waveforms}
\label{subsec:GW}
We compute approximate gravitational waveforms by evaluating the
quadrupole formula.  In the radiation zone, gravitational waves can be
described by a transverse-traceless, perturbed metric $h_{ij}^{TT}$
with respect to a flat spacetime. In the quadrupole formula,
$h_{ij}^{TT}$ is found from \citep{MTW}
\begin{equation}
h_{ij}^{TT}= \frac{2}{r} \frac{d^{2}}{d t^{2}} I_{ij}^{TT},
\label{eqn:wave1}
\end{equation}
where $r$ is the distance to the source, $I_{ij}$ the quadrupole
moment of the mass distribution (see Eq.~{[36.42b]} in
Ref.~\citep{MTW}), and where $TT$ denotes the transverse-traceless
projection.  Choosing the direction of the wave propagation to be
along the $z$-axis (rotational axis of the equilibrium star), we
determine the two polarization modes of gravitational waves from
\begin{equation}
h_{+}^{(z)} \equiv \frac{1}{2} (h_{xx}^{TT} - h_{yy}^{TT})
\mbox{~~~and~~~} 
h_{\times}^{(z)} \equiv h_{xy}^{TT}.
\end{equation}
For observers along the $z$-axis, we thus have
\begin{eqnarray}
\frac{r h_{+}^{(z)}}{M} &=& 
\frac{1}{2 M} \frac{d}{d t} (\dot{I}_{xx} - \dot{I}_{yy}),
\label{h+z}
\\
\frac{r h_{\times}^{(z)}}{M} &=& 
\frac{1}{M} \frac{d}{d t} \dot{I}_{xy}
\label{h-z}
.
\end{eqnarray}
Using the same procedure, the observers along the $x$-axis detect the
wave propagates as
\begin{eqnarray}
\frac{r h_{+}^{(x)}}{M} &=& 
\frac{1}{2 M} \frac{d}{d t} (\dot{I}_{yy} - \dot{I}_{zz}), \label{h+x}
\\
\frac{r h_{\times}^{(x)}}{M} &=& 
\frac{1}{M} \frac{d}{d t} \dot{I}_{yz} \label{h-x}
.
\end{eqnarray}
The number of time derivatives $I_{ij}$ that have to be taken out can
be reduced by using the continuity equation
(Eq.~[\ref{eqn:continuity}])
\begin{equation}
\dot{I}_{ij} = \int (\rho v^{i} x^{j} + \rho x^{i} v^{j}) d^{3}x,
\end{equation}
in equations (\ref{h+z}) -- (\ref{h-x}) (see Ref.~\citep{Finn}).

The spectrum of gravitational waveform can be computed as 
\begin{equation}
S^{(x,z)} =  |\tilde{h}_{+}^{(x,z)}|^2 + |\tilde{h}_{\times}^{(x,z)}|^{2},
\end{equation}
where 
\begin{equation}
\tilde{h}_{+, \times}^{(x,z)} = \int dt h_{+, \times}^{(x,z)} e^{i \omega t}.
\end{equation}

\subsection{Diagnostics}
We monitor the conservation of mass $M$ (Eq.~[\ref{eqn:M}]), angular
momentum $J$
\begin{equation}
J = \int \rho ( x v^{y} - y v^{x} ) d^3x,
\end{equation}
and the location of the center of mass $x^i_{\rm CM}$
\begin{equation}
x^{i}_{\rm CM} = \int \rho x^{i} d^3 x.
\end{equation}
Due to our flux-conserving difference scheme the mass $M$ is also
conserved up to a round-off error, except if matter leaves the
computational grid.

To monitor the development of the azimuthal modes ($m=1$, $2$, $3$,
$4$) and the one in the $z$-direction, we compute the following five
diagnostics
\begin{eqnarray}
D &=& \left< e^{i m \varphi} \right>_{m=1} 
\nonumber \\
&=&
  \frac{1}{M} \int \rho 
  \frac{x + i y}{\sqrt{x^{2}+y^{2}}} d^3 x
\label{eqn:dipole}
,\\
Q &=& \left< e^{i m \varphi} \right>_{m=2} 
\nonumber \\
&=&
  \frac{1}{M} \int \rho 
  \frac{(x^{2}-y^{2}) + i (2 x y)}{x^{2}+y^{2}} d^3 x
\label{eqn:quadrupole}
,\\
O &=& \left< e^{i m \varphi} \right>_{m=3} 
\nonumber \\
&=&
  \frac{1}{M} \int \rho 
  \frac{x (x^2 - 3 y^2) + i y (3 x^2 - y^2)}{(x^{2}+y^{2})^{3/2}} d^3 x
,\\
M_4 &=& \left< e^{i m \varphi} \right>_{m=4} 
\nonumber \\
&=&
  \frac{1}{M} \int \rho 
  \frac{(x^4 - 6 x^2 y^2 + y^4) 
    + i (4 x^2 y^2 (x^2 - y^2))}{(x^{2}+y^{2})^2} d^3 x,
\nonumber \\
&&\\
D_{z} &=&
  \frac{1}{M R_p} \int \rho |z| d^3 x,
\end{eqnarray}
where a bracket denotes the density weighted average.  When we compute
the four diagnostics in the equatorial plane ($D^{\rm (eq)}$, $Q^{\rm
  (eq)}$, $O^{\rm (eq)}$, $M_4^{\rm (eq)}$), we change the integral
volume from $d^3x$ to $dxdy$ and $M$ to $M_{\rm eq}$.  Note that
$M_{\rm eq}$ is the rest-mass density integrated only in the
equatorial plane.

\begin{figure*}
\centering
\includegraphics[keepaspectratio=true,width=12cm]{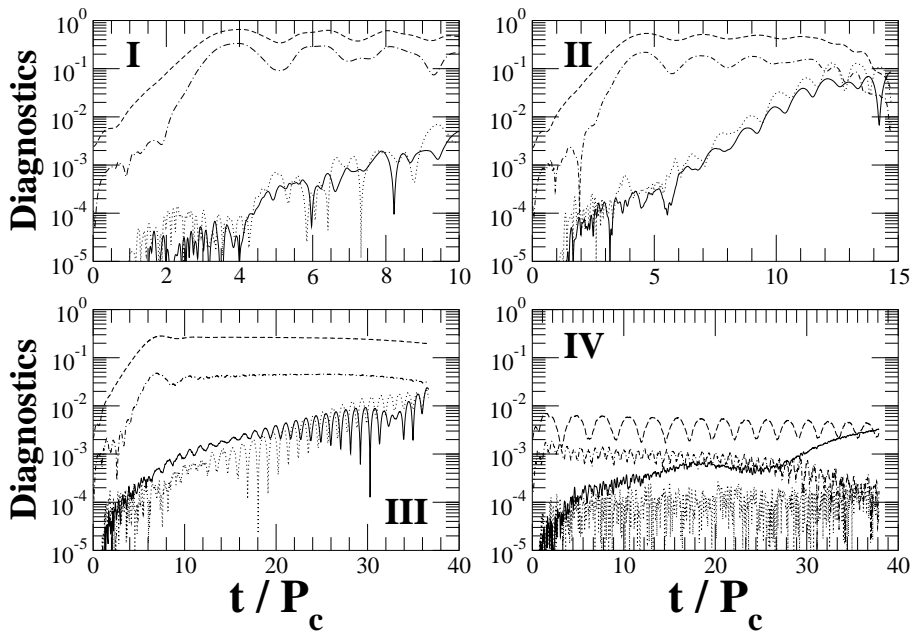}
\caption{
Diagnostics $|D|$, $|Q|$, $|O|$, $|M_4|$ as a function of $t/P_{\rm
  c}$ for four different rotating stars  (see Table
\ref{tab:equilibrium}).  Solid, dashed, dotted, and dash-dotted lines
denote $|D|$, $|Q|$, $|O|$ and $|M_4|$, respectively.  We terminate
our simulation when the relative error of the rest mass exceeds $\sim
0.01$\% for models I and II.  Hereafter $P_{\rm c}$ represents the
central rotation period at $t=0$.
}
\label{fig:lew_dig}
\end{figure*}

We also compute the spectra of the above five diagnostics as 
\begin{eqnarray}
|F_m|^2 &=& 
  \left| \frac{1}{M} \int dt \int d^3 x 
  \rho e^{i (\omega t - m\varphi)} \right|^{2} ~~~(m=1, \cdots, 4) ,
\nonumber \\
&& \\
|F_z|^2 &=& 
  \left| \frac{1}{M R_{\rm p}}\int dt \int d^3 x 
  |z| (\rho - \rho_{\rm avg}) e^{i \omega t} \right|^2.
\end{eqnarray}
Since $D_z$ does not oscillate around zero as we show later, we have
subtracted the time averaged density $\rho_{\rm avg}$ from the
original one to compute the spectrum.  Note also that we have only
integrated $D_z$ in time after its first global maximum.

\section{Numerical Results}
\label{sec:nr}
Here we show our evolution of the differentially rotating stars.  We
terminate the integration when the relative error of the rest mass
exceeds $\sim 10^{-4}$, since the only violation of the rest mass
conservation is caused by the matter outflow at the outer boundary of
computation.  We also terminate the integration when the time exceeds
$20 \sim 40$ central rotation periods, which are sufficient to enhance
all $m$ modes.  Note that our code never crashes throughout the
evolution.

To enhance any dynamically unstable mode, we disturb the initial
equilibrium density $\rho_{\rm eq}$ by a non-axisymmetric perturbation
according to \footnote{The numerical finite difference error is in
  principle sufficient to trigger instabilities, but starting from
  such a small amplitude it would take the instability significantly
  long to reach saturation.}
\begin{widetext}
\begin{equation}
\rho = \rho_{\rm eq}
\left[ 1 + 
  \delta^{(2)} \frac{x^{2} + 2 x y - y^{2}}{R_{\rm eq}^{2}} +
  \delta^{(4)} \frac{x^4 - 6 x^2 y^2 + y^4 + 4 x y (x^2 - y^2)}{R_{\rm
      eq}^{4}}
\right],
\label{eqn:DPerturb}
\end{equation}
\end{widetext}
where we set $\delta^{(2)} = \delta^{(4)} = 10^{-2}$.

We study four different differentially rotating stars, which are
detailed in Table \ref{tab:equilibrium} to investigate the nonlinear
behaviour of the non-axisymmetric dynamical bar instabilities.  We
choose the axis of rotation to align with the $z$ axis, and assume
planar symmetry across the equator.  We choose the Cartesian
coordinates with the computational grid points $401 \times 401 \times
101$ covering the equatorial diameter of the equilibrium star as $121$
grid points.

\begin{table}[htbp]
\begin{center}
\leavevmode
\caption{
Four different rotating equilibrium stars in Newtonian gravity of
$\Gamma = 2$, $d/R_{\rm eq}=1$.
}
\begin{tabular}{c c c c}
\hline
\hline
Model & $R_{\rm p} / R_{\rm eq}$ & $T/W$ & $V_{\rm Ntw}$
\\
\hline
I & $0.225$ & $0.281$ & $8.29 \times 10^{-5}$
\\
II & $0.250$ & $0.277$ & $8.79 \times 10^{-5}$
\\
III & $0.275$ & $0.268$ & $7.95 \times 10^{-5}$
\\
IV & $0.300$ & $0.256$ & $9.47 \times 10^{-5}$
\\
\hline
\hline
\end{tabular}
\label{tab:equilibrium}
\end{center}
\end{table}

\begin{figure*}
\centering
\includegraphics[keepaspectratio=true,width=12cm]{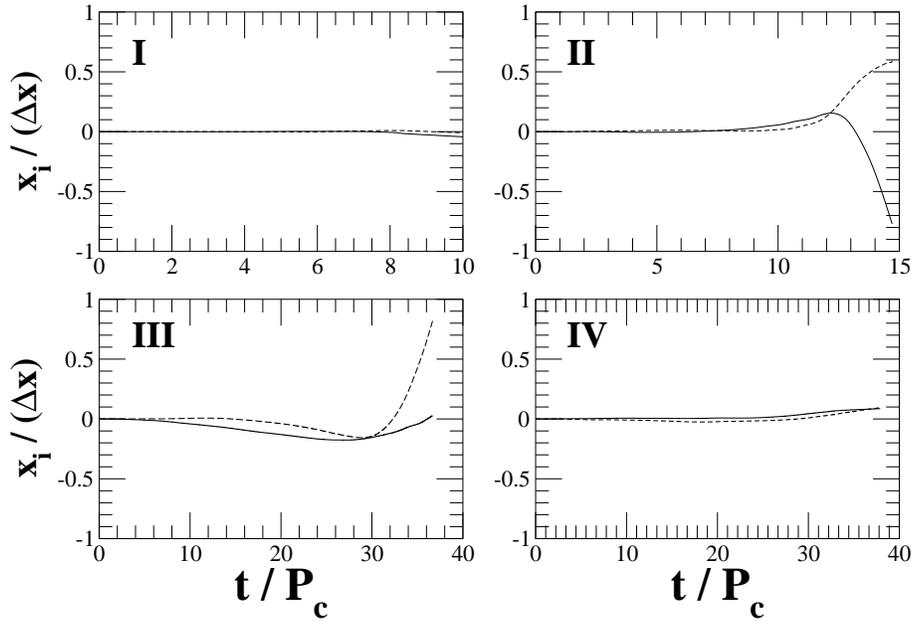}
\caption{
Center of mass as a function of $t/P_{\rm c}$.  Solid, dashed line
denotes the one of $x$ and $y$ direction, respectively.
}
\label{fig:lew_cm}
\end{figure*}

\begin{figure*}
\centering
\includegraphics[keepaspectratio=true,width=12cm]{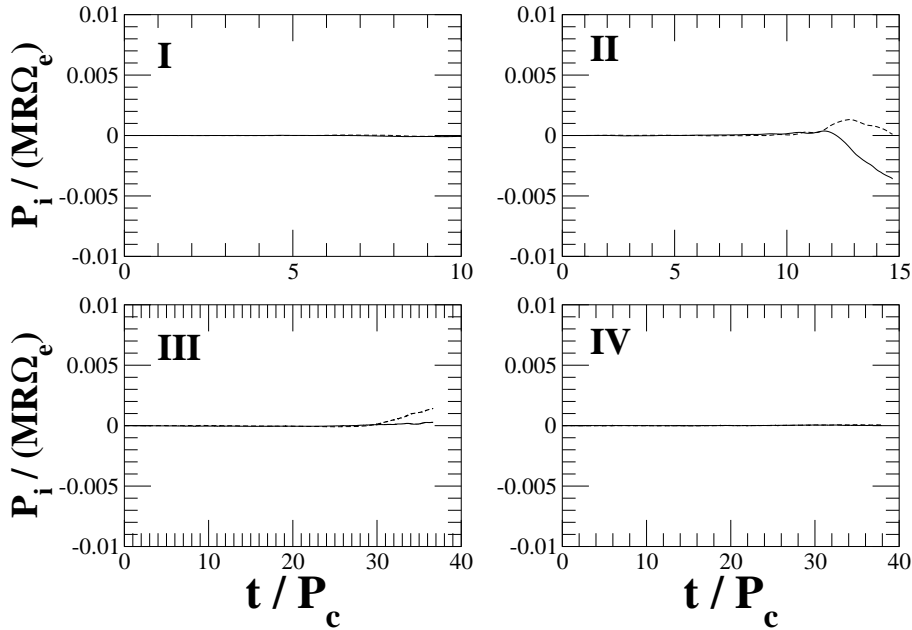}
\caption{
Linear momentum as a function of $t/P_{\rm c}$.  Solid and dashed line
denotes $x$ and $y$ component of the linear momentum, respectively.
$M$ and $\Omega_e$ represent the total rest-mass and the angular
velocity at the equatorial surface at $t = 0$, respectively.
}
\label{fig:lew_lm}
\end{figure*}

\begin{figure*}
\centering
\includegraphics[keepaspectratio=true,width=12cm]{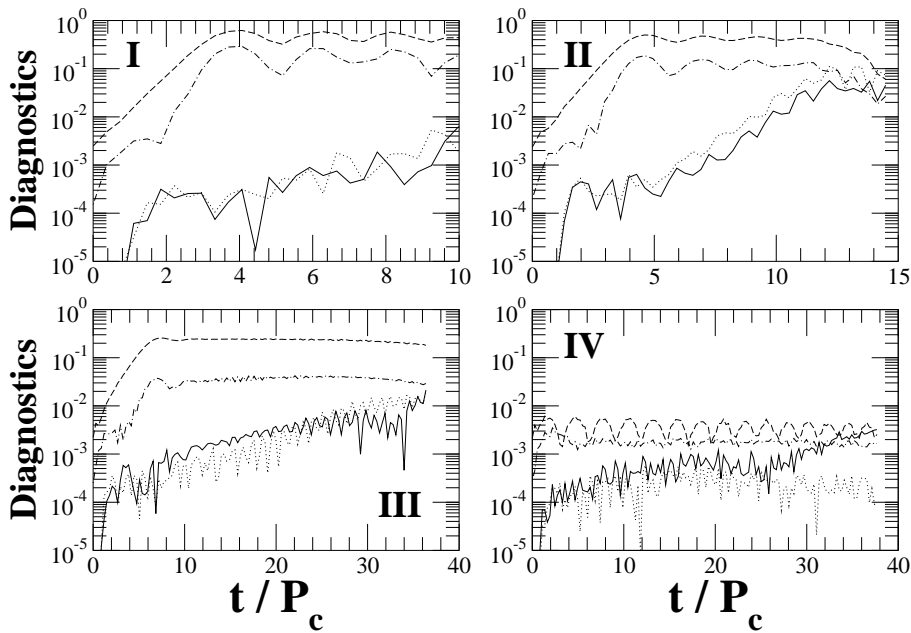}
\caption{
Same as Fig.~\ref{fig:lew_dig}, but the diagnostics are only computed
in the equatorial plane.
}
\label{fig:lewdig_eq}
\end{figure*}

We show the amplitudes of our four diagnostics for all four models in
Fig.~\ref{fig:lew_dig}.  At the first stage of evolution, the $m=2$
diagnostic grows exponentially in models I, II, and III, while it
stays around the amplitude of $t=0$ in model IV.  Therefore the star
of models I, II, and III is determined as dynamically unstable against
bar mode, while that of model IV is stable.  For the dynamically bar
unstable stars (models I, II, and III), the $m=2$ diagnostic grows
exponentially but the other remaining $m$ modes do not grow at the
first evolution stage when imposing a small perturbation
(Fig.~\ref{fig:lew_dig}).  This result is consistent to the linear
perturbation analysis of the dynamically bar unstable stars, which
shows that the only dynamically unstable $m$ mode is $m=2$.  Also the
result confirms us that the amplitude of perturbation at $t=0$ is
adequate to treat the system linearly ($\delta^{(2)} \approx
10^{-2}$).  After that stage the $m=4$ diagnostic grows exponentially
because of the secondary harmonic of $m=2$ mode, and then the odd $m$
modes are also enhanced.  The odd $m$ modes are excited even if we do
not impose the perturbation of their corresponding modes in the
equilibrium star, since the finite differencing scheme always
generates a small amount of all $m$ modes (Fig.~\ref{fig:lew_dig}).
However a small fluctuation at the wavefront should occur in nature so
that the existence of all $m$ modes, when the bar forms, are quite
natural in reality.

\begin{figure*}
\centering
\includegraphics[keepaspectratio=true,width=12cm]{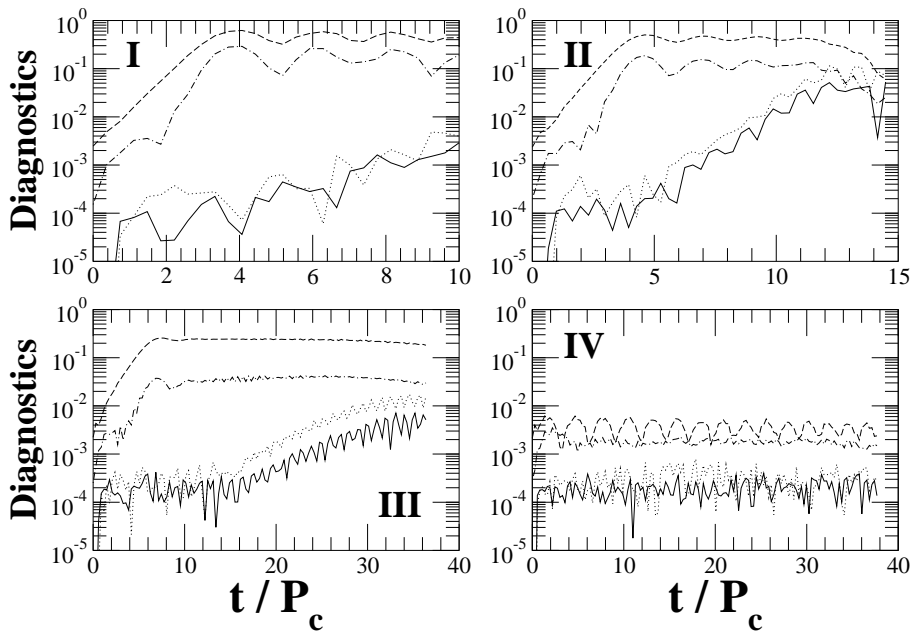}
\caption{
Same as Fig.~\ref{fig:lewdig_eq}, but the center of mass is adjusted
to zero in every snapshots. 
}
\label{fig:lewdig_ea}
\end{figure*}

We have monitored the center of mass and the linear momentum
throughout the evolution to guarantee that we do not impose any
additional physics in the system.  Fig.~\ref{fig:lew_cm} shows the
center of mass of the four different stars throughout the evolution.
We have confirmed that the numerical error only allows the star to
change the center of mass within the one computational grid.  We have
also checked the linear momentum conservation in
Fig.~\ref{fig:lew_lm}, which shows that the relative error is less
than 1\% of the total value constructed by the total mass and the
velocity at the equilibrium equatorial surface.  In order to check
whether the center of mass condition significantly affects the
diagnostics, we have also computed the following two types of
diagnostics in the equatorial plane (e.g. \citep{CQF07}).  One is the
diagnostic with the same coordinate as in the simulation, while the
other is the one with the coordinate where the center of mass is
adjusted to zero in every snapshots.  Since the equatorial diagnostic
(Fig. \ref{fig:lewdig_eq}) reproduces all characteristics of the one
obtained from the three-dimensional computation
(Fig.~\ref{fig:lew_dig}), the equatorial diagnostic may represent the
three-dimensional one.  We compare Figs.~\ref{fig:lewdig_eq} and
\ref{fig:lewdig_ea},  to focus on the effect of the center of mass
condition on the diagnostics.  For models I and II, the adjustment of
the center of mass reduces the amplitude of $D^{\rm (eq)}$ and $O^{\rm
  (eq)}$for $t \lesssim 5 P_{c}$.  However the condition does not
change the exponential growth of $D^{\rm (eq)}$ and $O^{\rm (eq)}$
after $t \gtrsim 5 P_{c}$.  For models III and IV the amplitude of
$D^{\rm (eq)}$ has been reduced so that the system is stable to
$m=1$.  Therefore the linear growth of $D^{\rm (eq)}$ in models III
and IV is the outcome of the violation of the center of mass
condition.

\begin{figure*}
\centering
\includegraphics[keepaspectratio=true,width=16cm]{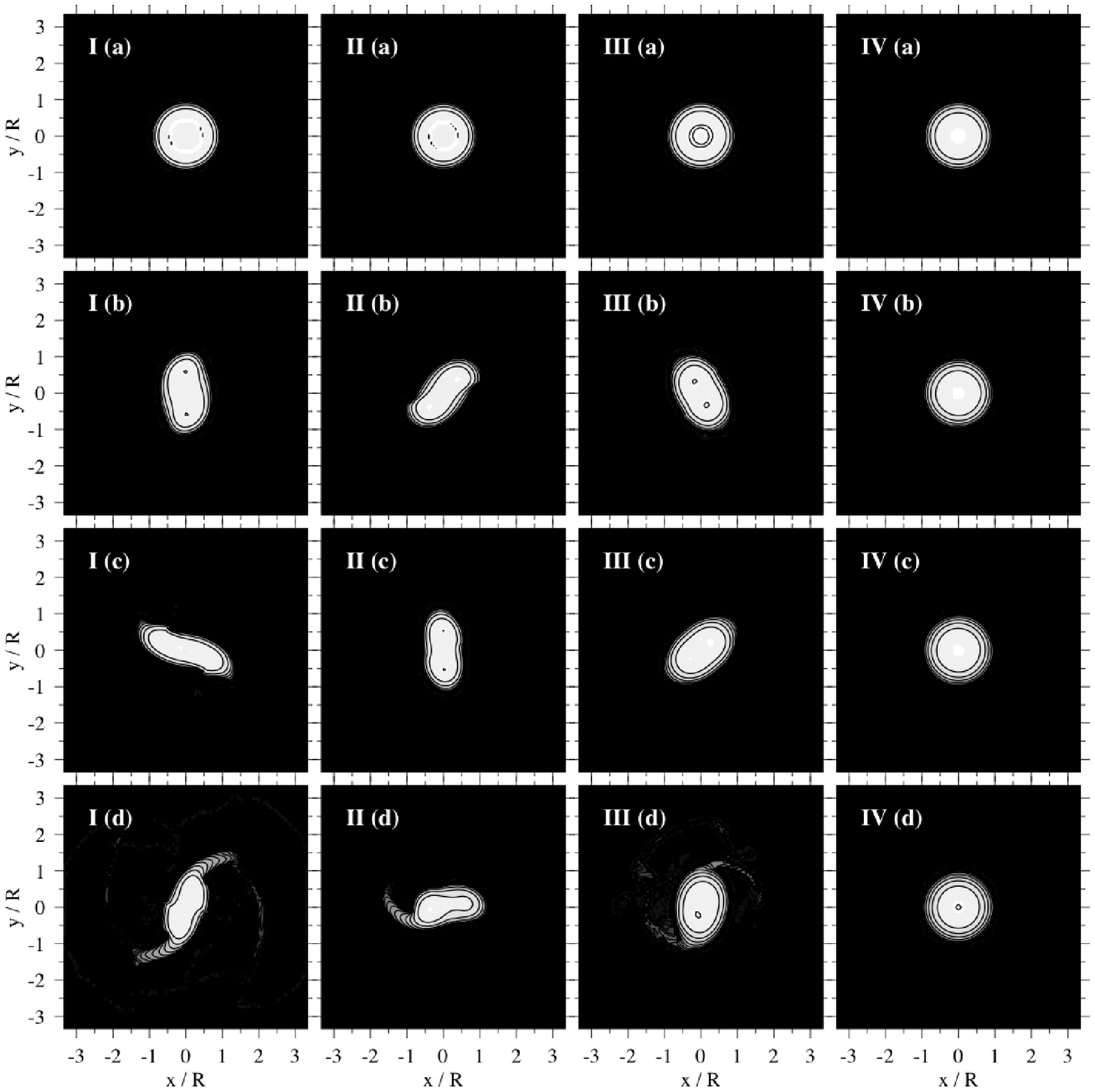}
\caption{
Density contours in the equatorial plane throughout the evolution.
Snapshots are plotted at
($t / P_{\rm c}, \rho_{\rm max} / \rho_{\rm max}^{(0)}$, d) $=$ 
I(a) ($3.70 \times 10^{-4}$, $1.00$, $0.25000$), 
I(b) ($2.40$, $1.21$, $0.25000$), 
I(c) ($5.92$, $1.27$, $0.25000$), 
I(d) ($8.87$, $1.46$, $0.25000$), 
II(a) ($3.29 \times 10^{-4}$, $1.00$, $0.20625$), 
II(b) ($3.95$, $1.20$, $0.20625$), 
II(c) ($7.90$, $1.29$, $0.20625$), 
II(d) ($11.85$, $1.58$, $0.20625$), 
III(a) ($2.98 \times 10^{-4}$, $1.00$, $0.25000$), 
III(b) ($11.93$, $1.13$, $0.25000$), 
III(c) ($23.85$, $1.19$, $0.25000$), 
III(d) ($35.78$, $1.27$, $0.25000$), 
IV(a) ($2.75 \times 10^{-4}$, $1.00$, $0.25000$), 
IV(b) ($12.10$, $1.05$, $0.25000$), 
IV(c) ($24.20$, $1.10$, $0.25000$), 
IV(d) ($36.30$, $1.16$, $0.25000$), 
where $\rho_{\rm max}$ is the maximum rest mass density and $\rho_{\rm
  max}^{(0)}$ is the maximum rest mass density at $t=0$.  The contour
line denotes $\rho/\rho_{\rm max} = 10^{(16 - i)d} (i = 1, \ldots,
15)$.  Hereafter $R$ denotes the equatorial radius at $t=0$.
}
\label{fig:lewxy_con}
\end{figure*}

\begin{figure*}
\centering
\includegraphics[keepaspectratio=true,width=16cm]{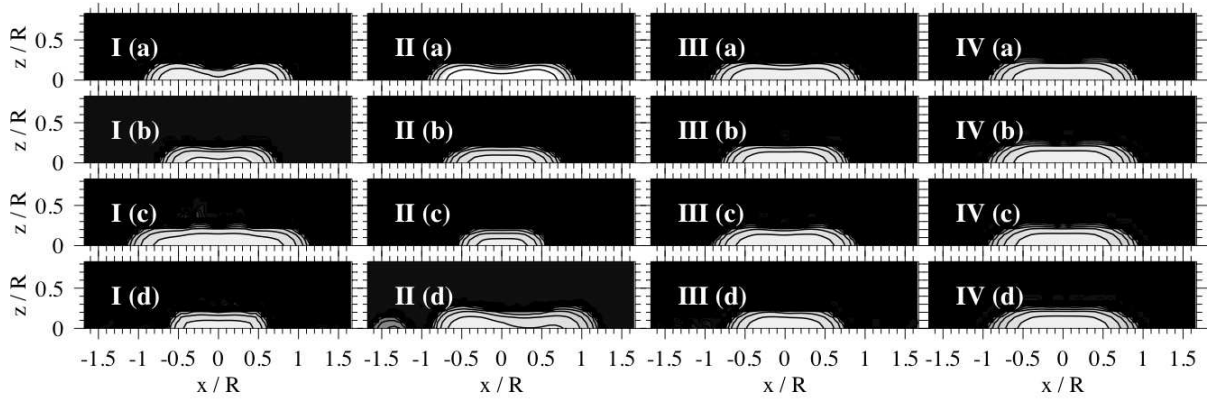}
\caption{
Same as Fig. \ref{fig:lewxy_con} but in the meridional plane.
}
\label{fig:lewxz_con}
\end{figure*}

We also show our equatorial and the meridional density snapshots
throughout our integration in Figs. \ref{fig:lewxy_con} and
\ref{fig:lewxz_con}.  The symmetry breaking of the bar structure
occurs clearly at the time when the spiral arm forms in the equatorial
snapshot and in the meridional plane.  This becomes clear when we
focus on the final snapshots of models I, II, and III.

\begin{figure*}
\centering
\includegraphics[keepaspectratio=true,width=16cm]{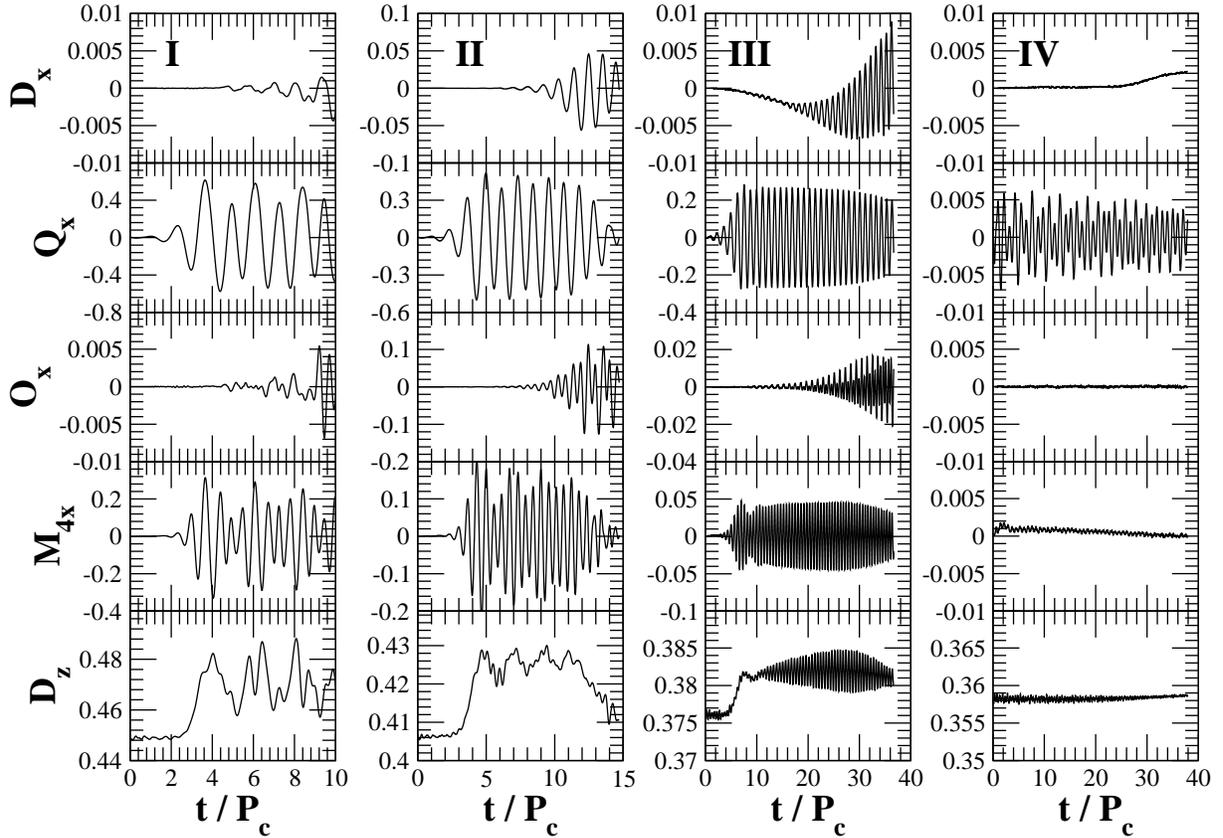}
\caption{
Diagnostics $\Re [D]$, $\Re [Q]$, $\Re [O]$, $\Re [M_4]$, and $D_z$ as
a function of $t/P_{\rm c}$ for four different rotating stars  (see
Table \ref{tab:equilibrium}).
}
\label{fig:lew_dx}
\end{figure*}

\begin{figure*}
\centering
\includegraphics[keepaspectratio=true,width=16cm]{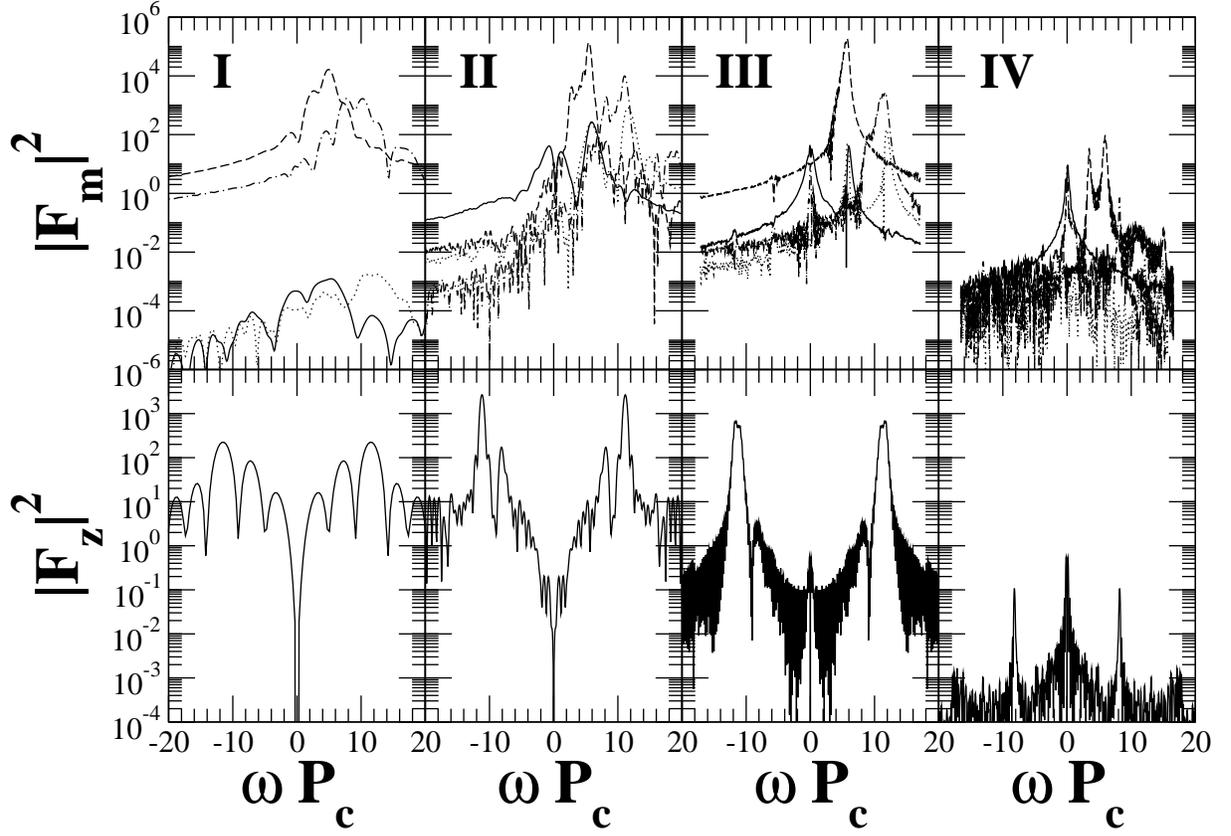}
\caption{
Spectra $|F_m|^2$ and $|F_z|^2$ as a function of $\omega P_{\rm c}$
for four different rotating stars (see Table
\ref{tab:equilibrium}).  Solid, dashed, dotted, and dash-dotted line
of $|F_m|^2$ denote the values of $m=1$, $2$, $3$, and $4$,
respectively.
}
\label{fig:lew_spk}
\end{figure*}

We also in Fig.~\ref{fig:lew_dx} show the diagnostics which contain
both amplitude and phase.  In order to make the picture clear, we
first concentrate on the model III, the weakest dynamically unstable
bar system of three models.

The behaviours in the diagnostics are clearly understood once we
compute the spectra of the diagnostics (Fig.~\ref{fig:lew_spk}).  From
the spectra we find the following two remarkable issues.  One is that
the spectra $|F_1|^2$,  $|F_2|^2$,  $|F_3|^2$ take a peak around
$\omega_{\rm bar} \approx 5 \sim 6 P_c^{-1}$ for models I, II, III,
and the other is that $|F_3|^2$,  $|F_4|^2$, $|F_z|^2$ take a peak
around $\omega_{\rm quad} \approx 2 \omega_{\rm   bar} \approx 10$ --
$12 P_c^{-1}$for bar unstable stars.  Combining the present feature
with the behaviour of the five  diagnostics explained before
(Fig.~\ref{fig:lew_dig}), the dynamically unstable bar acts as
follows.

Firstly the $m=2$ mode grows and acts as a dominant mode of all
because of the dynamical bar instability.  Next the $m=4$ mode grows
because of the secondary harmonic of the $m=2$ mode.  In fact the
saturation amplitude of the $m=4$ is approximately $\approx 0.3$ for
model I, $0.2$ for model II, and $0.04$ for model III, all of which
are the order of the square of the saturation amplitude of the $m=2$
($\approx 0.6^2$ for model I, $0.5^2$ for model II, $0.2^2$ for model
III).  After that Faraday resonance occurs, which is clearly found in
both $D_z$ and $|F_z|^2$ from the fact $\omega_{\rm quad} \approx 2
\omega_{\rm bar}$.

Note that Faraday resonance occurs in the fluid mechanics when the
oscillation of the vertical direction is twice ($2\omega$) as much as
the one in the horizontal direction ($\omega$) in the weakly nonlinear
interaction \citep{DR81,MH90}.  The reason why the resonance does not
clearly appear in model I is either the strongly nonlinear effect or
the insufficient duration time of quasi-periodic oscillation for
computing the spectrum.  Then, there is a resonance between $m=1$ and
$m=2$, $m=3$ and $m=4$.  The possibility of such resonances is three
wave interaction: either $m=1$ ($\omega_{\rm bar}$) and $m=2$
($\omega_{\rm bar}$) generates $m=3$ ($\omega_{\rm bar} + \omega_{\rm
  bar}$) or $m=3$ ($2 \omega_{\rm bar}$) and $m=2$ ($\omega_{\rm
  bar}$) generates $m=1$ ($2 \omega_{\rm bar} - \omega_{\rm bar}$) in
the dominant part.  It is the fact found in the nonlinear behaviour of
the dynamically unstable bar system.

\begin{figure*}
\centering
\includegraphics[keepaspectratio=true,width=12cm]{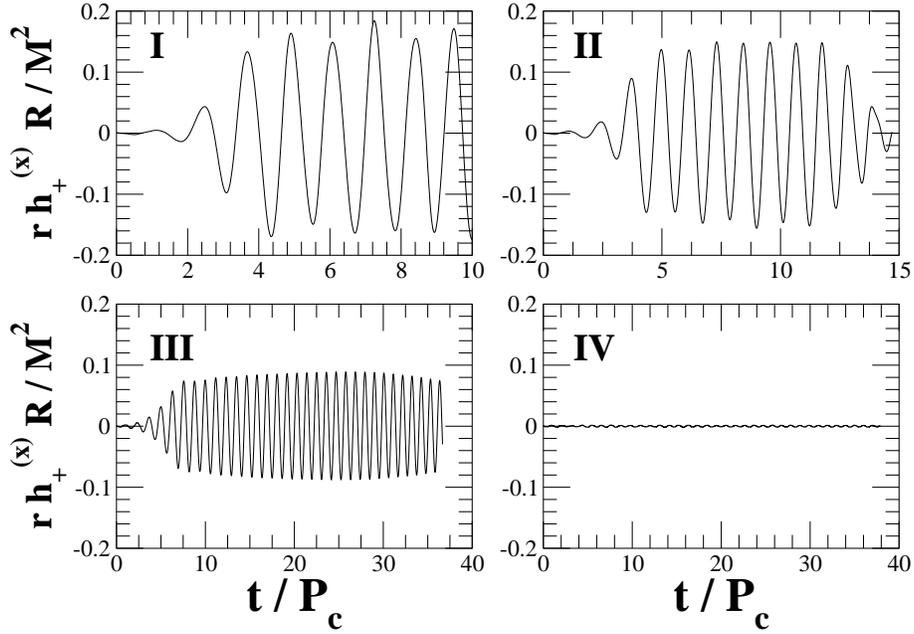}
\caption{
Gravitational waveform ($+$ mode) observed at the $x$ axes for four
different rotating stars (See Table \ref{tab:equilibrium}).  Note that
$\times$ mode is identically zero because we adopt the equatorial
symmetry.
}
\label{fig:lewx_gw}
\end{figure*}

\begin{figure*}
\centering
\includegraphics[keepaspectratio=true,width=12cm]{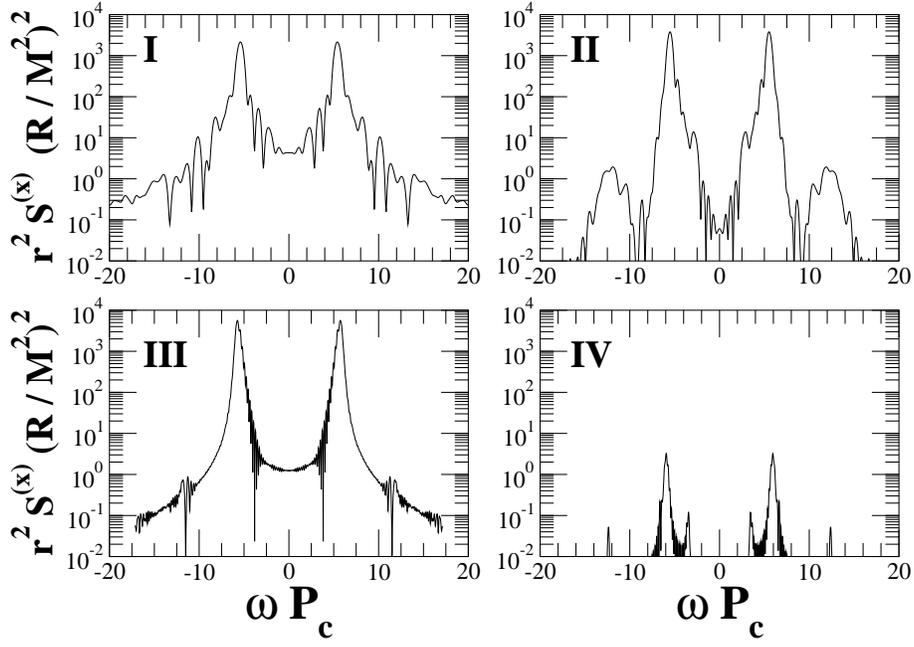}
\caption{
Spectra of gravitational waveform ($+$ mode) observed at the $x$ axes
for four different rotating stars (See Table \ref{tab:equilibrium}).
}
\label{fig:lewx_spk}
\end{figure*}

\begin{figure*}
\centering
\includegraphics[keepaspectratio=true,width=12cm]{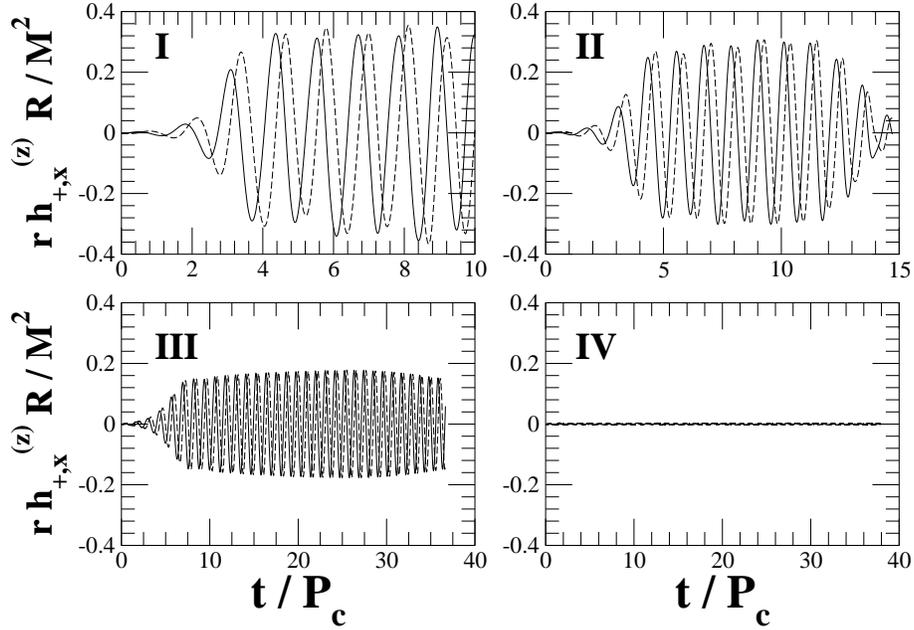}
\caption{
Gravitational waveform observed at the $z$ axes for four different
rotating stars (See Table \ref{tab:equilibrium}).  Solid and dashed
line denotes $+$ mode and $\times$ mode, respectively.
}
\label{fig:lewz_gw}
\end{figure*}

\begin{figure*}
\centering
\includegraphics[keepaspectratio=true,width=12cm]{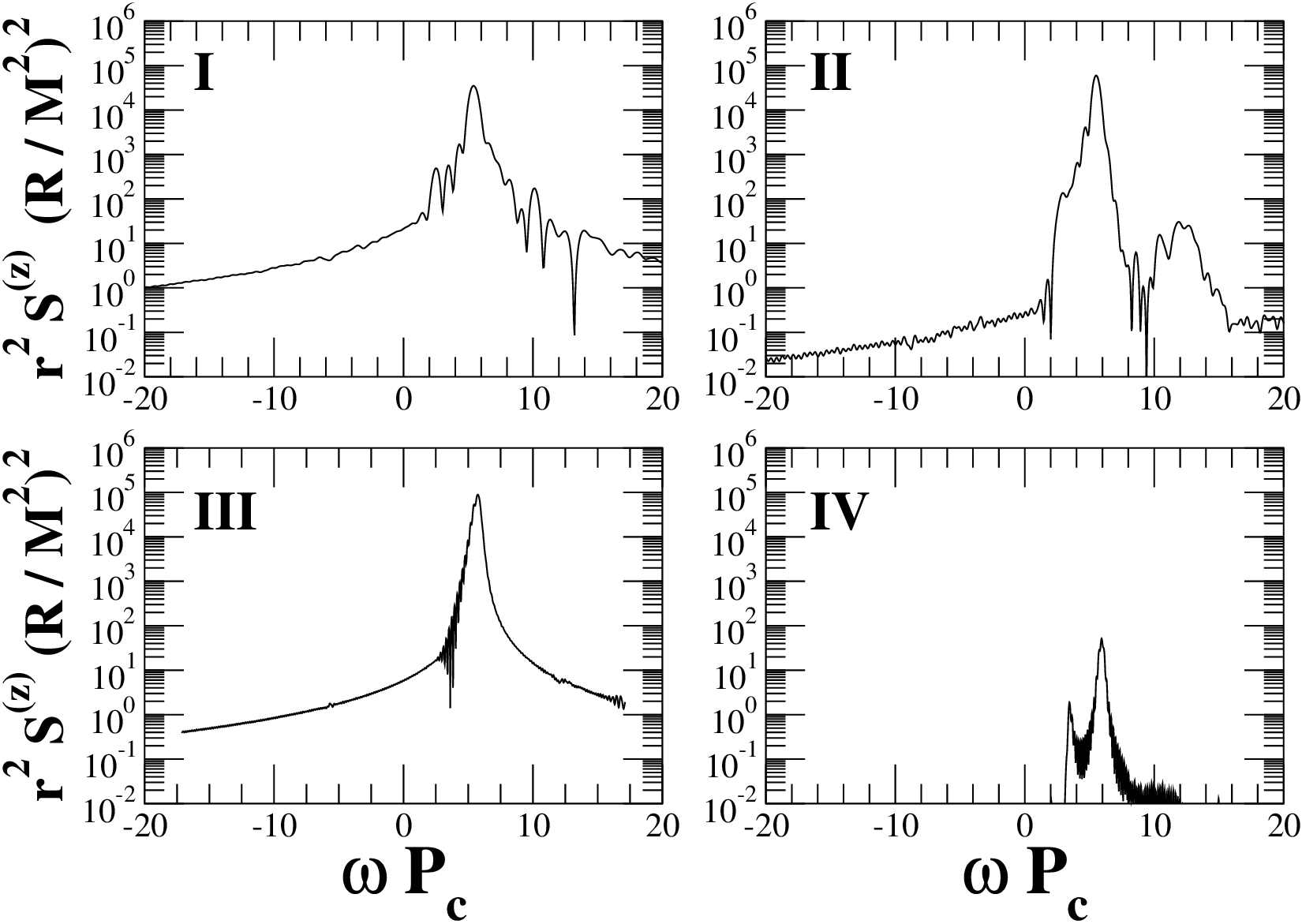}
\caption{
Spectra of gravitational waveform observed at the $z$ axes for four
different rotating stars (See Table \ref{tab:equilibrium}).
}
\label{fig:lewz_spk}
\end{figure*}

The gravitational waveform and its spectrum have been computed by the
quadrupole formula observed along the rotational axis and in the
equatorial plane (Figs.~\ref{fig:lewx_gw} --
~\ref{fig:lewz_spk}).  There are two remarkable features in
gravitational waves from the viewpoint of nonlinear behaviour.  One is
that the quasi-periodic oscillation does not last until the radiation
reaction timescale but decays because of the symmetry breaking of the
dynamical bar.  The duration period is related to the degree of
nonlinearity of the bar mode instability, which is estimated from the
inclination angle of the amplitude of the $m=1$ (Re[$D$]) and $m=3$
(Re[$O$]) diagnostics.  In the present case, the duration period of
the bar structure is estimated as $\sim 10 P_c$ for model I, $\sim 15
P_c$ for model II, and $\sim 35 P_c$ for model III.  The other is that
Faraday resonance has clearly appeared in the spectrum of
gravitational waveform observed at least in the equatorial plane.
Since we adopt quadrupole formula to compute gravitational waves, the
higher order harmonics of the unstable bar mode such as $m=4$ mode
cannot be seen in this spectrum.  Therefore a peak around $\omega
\approx 12 P_c^{-1}$ in Fig.~\ref{fig:lewx_spk} indicates the fact of
an oscillation along the $z$-axis, which is the evidence of {\it
  Faraday resonance}.  We have also computed the gravitational
waveform and its spectrum observed along the $z$-axis and found that
there is no peak around $\omega \approx 12 P_c^{-1}$ in model III
(Fig.~\ref{fig:lewz_spk}).  The fact also supports that a peak around
$\omega \approx 12 P_c^{-1}$ in Fig.~\ref{fig:lewx_spk} is the outcome
of Faraday resonance, since an oscillation along $z$-direction can be
clearly observed by gravitational waves in the equatorial plane, not
in the rotation axis.  When we increase the degree of nonlinearity,
the above feature of the Faraday resonance in gravitational waves can
be also seen in the equatorial plane.

\section{Conclusion}
\label{sec:Conclusion}
We investigate the nonlinear effects of dynamically bar unstable stars
by means of three dimensional hydrodynamic simulations in Newtonian
gravity. In order to follow the bar shape as long as possible, the
initial amplitudes for odd azimuthal perturbations are significantly
suppressed in our models.  

We find interesting mode coupling in the dynamically unstable system
in the nonlinear regime, and that only before the destruction of the
bar.  The quasi-periodic oscillation mainly along the rotational axis
is induced. The characteristic frequency is twice as big as that of
the dynamically unstable bar mode.  This feature is quite analogous to
the Faraday resonance.  Although our finding is only supported by the
weakly nonlinear theory of fluid mechanics, we have also found the
same feature of parametric resonance even in the strongly nonlinear
regime.  There is one qualitative difference between Faraday resonance
and our numerical result.  Faraday resonance has lower frequency than
that of the forced oscillation, while our result has higher frequency
than that of the bar unstable mode.  The fact can be understood by the
different regime of the media.  Since the media of the rotating star
is a perfect fluid, which is only contained inside the star, there
should be a cutoff frequency to be amplified.  In fact, introducing a
cutoff frequency with a polar radius of the star and the sound speed
computed by the mean density, the cutoff frequency $\omega_{\rm cut}$
is estimated as $\omega_{\rm cut} \approx \omega_{\rm bar}$.
Therefore a higher frequency than the bar is amplified.  The fact also
indicates that our finding is interpreted as a parametric resonance.

We also discuss the detectability of Faraday resonance by
gravitational waves.  The characteristic frequency and the amplitude
of gravitational waves can be estimated as
\begin{widetext}
\begin{eqnarray}
f_{\rm bar} &\sim& 
  2 \left( \frac{10 {\rm km}}{R} \right) 
  \left( \frac{T/W}{0.25} \right)^{1/2} 
  \left( \frac{M/R}{0.15} \right)^{1/2} 
  [{\rm kHz}]
,\\
h_{\rm bar} &\sim& 
  2 \times 10^{-23} 
  \left( \frac{M}{1.4 M_{\odot}} \right)
  \left( \frac{20 {\rm M pc}}{d_{\rm obs}} \right)
  \left( \frac{M/R}{0.15} \right)
  \left( \frac{T/W}{0.25} \right)
,
\end{eqnarray}
\end{widetext}
where $d_{\rm obs}$ is the distance from the observer.  If the bar
formation occurs in Virgo cluster, quasi-periodic waveform can be
detected in the second generation of gravitational wave detectors such
as Advanced LIGO, Advanced Virgo, and Large-scale Cryogenic
Gravitational wave Telescope (LCGT), or in the third generation
European Gravitational Wave Observatory.  The frequency of the
parametric resonance is around twice as big of that of the bar
unstable mode, and the amplitude of the parametric resonance is
roughly two orders lower ($\approx 1\%$) then that of bar unstable
mode.  The detection of gravitational waves from parametric resonance
may explore the nonlinear phase of the dynamically bar unstable stars
such as determining the saturation amplitude of gravitational waveform
from the bar unstable system, parametric resonance, and the duration
period of the bar structure.

\acknowledgments
MS thanks Nils Andersson, Silvano Bonazzola, Brandon Carter, Pablo
Cerda-Duran, Eric Gourgoulhon, Kei Kotake, Ewald M\"uller, Luciano
Rezzolla, Shin Yoshida for discussion.  MS also thanks Misao Sasaki
for his kind hospitality at the Yukawa Institute for Theoretical
Physics, where part of this work was done. This work was supported in
part by the STFC rolling grant (No.~PP/E001025/1), by the PPARC grant
(No.~PPA/G/S/2002/00531) at the University of Southampton, and by the
Grant-in-Aid for the 21st Century Center of Excellence in Physics at
Kyoto University.  Numerical computations were performed on the
myrinet nodes of Iridis compute cluster in the University of
Southampton, and on the SGI-Altix3700 in the Yukawa Institute for
Theoretical Physics, Kyoto University.


\end{document}